\documentstyle[aps,epsf]{revtex}

\newcommand{\expect}[1]{\langle{#1}\rangle}

\newcommand{\bramket}[3]{\langle\,{#1}\,|\,{#2}\,
            |\,{#3}\,\rangle}
\newcommand{\bqn}{\begin{eqnarray}}
\newcommand{\eqn}{\end{eqnarray}}
\newcommand{\beq}{\begin{equation}}
\newcommand{\eeq}{\end{equation}}
\newcommand{\x}{\times}

\newcommand{\us}{\mbox{\boldmath$\sigma$}}
\newcommand{\ut}{\mbox{\boldmath$\tau$}}
\newcommand{\ovl}[1]{\overline{#1}}
\newcommand{\mbf}[1]{\mbox{\boldmath$#1$}}
\newcommand{\CL}{{\cal L}}
\newcommand{\CO}{{\cal O}}

\begin{document}

\title{\bf Effective $T$-odd $P$-even  hadronic interactions from
  quark models}

\author{ Michael Beyer}
\address{Max Planck AG ``Vielteilchentheorie'', Rostock University, 18051
Rostock, Germany}

\maketitle

\begin{abstract}
  Tests of time reversal symmetry at low and medium energies may be
  analyzed in the framework of effective hadronic interactions. Here,
  we consider the quark structure of hadrons to make a connection to
  the more fundamental degrees of freedom. It turns out that for
  $P$--even $T$--odd interactions hadronic matrix elements evaluated
  in terms of quark models give rise to factors of 2 to 5. Also, it is
  possible to relate the strength of the anomalous part of the
  effective $\rho$ type $T$--odd $P$--even tensor coupling to quark
  structure effects.
\end{abstract}

\vspace{1ex}


\section{Introduction} 

First evidence of the violation of time reversal symmetry has been
found in the Kaon system~\cite{kaon}.  Despite strong efforts no other
signal of violation of time reversal symmetry has been found to date.
However, by now, studying time reversal symmetry has become a corner
stone of the search for physics beyond the Standard Model of
elementary particles~\cite{wolf}.  Some alternatives or extensions of
the Standard Model are due to dynamical symmetry breaking, Multi Higgs
models, spontaneous symmetry breaking, grand unified theories (e.g.
SO(10)), extended gauge groups (leading e.g. to right-handed bosons
$W_R$ in left-right symmetric models), Super Symmetric (SUSY)
theories, etc., each implying specific ways of $CP$ violation. For a
recent review of models relevant in the context of $CP$ violation see
e.g.~\cite{her95}, and refs. therein.

These theories ``beyond'' the standard model are formulated in terms
of quarks and leptons whereas nuclear low energy tests of $CP$ involve
hadronic degrees of freedom (mesons and
nucleons)~\cite{boehm95,gould94}.  To extract hadronic degrees of
freedom from observables one may introduce effective $T$--odd nucleon
nucleon potentials~\cite{her66}, or more specific $T$--odd mesonic
exchange potentials~\cite{bry71,sud68,sim75,sim77,hax83,her87}.  As in
the context of $P$-violation see e.g.~\cite{ade85}, these potentials
have been proven quite useful to treat the nuclear structure part
involved and to extract effective $T$--odd hadronic coupling
constants~\cite{tow94,gud93,bey89,bey93,hax93,mus94}. In turn they
allow to compare the sensitivity of different experiments, which has
been done recently in ref.~\cite{her95}.  However, in order to compare
upper bounds on a more fundamental level of $T$--odd interactions, it
is necessary to relate hadronic degrees of freedom to quark degrees of
freedom in some way. This step is hampered by the absence of a
complete solution of quantum chromo dynamics (QCD) at the energies
considered here. In many cases a rough estimate in the context of time
reversal violation may be sufficient, and, in the simplest case,
factors arising from hadronic structure may be neglected.  In the
context of $P$--odd time reversal violation e.g. concepts such as PCAC
and current algebra~\cite{cre79} have been utilized to improve the
evaluation of hadronic structure effects. In the $P$--even case, which
is considered here, this approach is not applicable (no Goldstone
bosons involved here). However,
it may be useful to utilize quark models
specifically designed for and quite successful in describing the low
energy sector.

In fact, experimental precision tests still continue to make progress
and so theorists face a renewed challenge to translate these
experimental constrains to a more fundamental interaction level.  The
purpose of the present paper is to give estimates on hadronic matrix
elements that arise when relating quark operators to the effective
hadronic parameterizations of the $P$--even $T$--odd interaction. These
are the charge $\rho$ type exchange and the axial vector type exchange
nucleon nucleon interaction~\cite{sud68,sim75,sim77}. They will
shortly be outlined in the next section. The ansatz to calculate $NN$
matrix elements from the quark structure is described in section III.
The last section gives the result for different types of quark models
and a conclusion.
 
For completeness, note that in general also $T$-odd and $P$-odd
interactions are possible, and in fact most of the simple extensions
of the standard model mentioned above give rise to such type of
$T$--violation.  Parameterized as one boson exchanges they lead e.g. to
effective pion exchange potentials that are essentially long range,
see~\cite{her95}. Limits on $P$--odd $T$--odd interactions are rather
strongly bound by electric dipole moment measurements, in particular
by that of the neutron~\cite{alt92,smi90,pen92,mon94}. In contrast
bounds on $P$--even $T$--odd interactions are rather weak.  Note, also
that despite theoretical considerations~\cite{her95a,eng96} new experiments
testing generic $T$--odd $P$--even observables have been suggested;
for the present status  see e.g. refs.~\cite{boehm95,gould94}.

\section{Effective $T$-odd $P$-even nucleon nucleon interactions}

Due to the moderate energies involved in nuclear physics tests of time
reversal symmetry, hadronic degrees of freedom are useful and may be
reasonable to analyze and to compare different types of experiments.
For a recent discussion see ref.~\cite{her95}. In the following only
$T$-odd and $P$-{\em even} interactions will be considered. They may
be parameterized in terms of effective one boson exchange potentials.
Due to the behavior under $C$, $P$, and $T$ symmetry transformations,
see e.g.~\cite{zuber}, two basic contributions are possible then: a
charged $\rho$ type exchange~\cite{sim75,sim77} and an axial vector
exchange~\cite{sud68}.  The effective $\rho$ type $T$--odd interaction
is $C$--odd due to the phase appearing in the isospin sector and is
only possible for charged $\rho$ exchange. It has
been suggested by Simonius and Wyler, who used the tensor part to
parameterize the interaction~\cite{sim77},
\begin{equation}
{\cal L}^T_{\rho NN} =  g^T_{\rho NN}
\bar N\sigma_{\mu\nu} 
\frac{(p_{f}-p_{i})^\nu}{2m_N}
\frac{1}{\sqrt{2}}\left(\rho^{+\mu}\tau^- - \rho^{-\mu}\tau^+\right)N.
\label{eqn:VNN}
\end{equation} 
There is some question of whether to choose an ``anomalous''
coupling~\cite{her95,herINT}, viz.  $g^T_{\rho NN}=\kappa \tilde g^T_{\rho
  NN}$.  The numerical value of $\kappa$ is usually taken to be $3.7$
close to the strong interaction case~\cite{sim77}.  We shall see in the
following that it is not unreasonable to introduce such a factor since
in may be related to ``nucleonic structure effects'', which are not of $T$
violating origin (similar to nuclear structure effects that are also
treated separately).  
Combining the $T$--odd vertex with the appropriate
$T$--even vertex leads to the following effective $T$--odd $P$--even
one boson exchange $NN$ interaction,
\begin{equation}
V_\rho^T=i\;\frac{g^T_{\rho NN}g_{\rho NN}}
          {8m^2_N({\bf q}^2+m_\rho^2)}\;
          (\us_1-\us_2)\cdot{\bf q}\x{\bf p}\quad
          (\ut_1\x\ut_2)_0,
\end{equation}
where   ${\bf q}={\bf p}_f - {\bf p}_i$, and ${\bf p}=({\bf p}_f+{\bf
  p}_i)/2$, and $g_{\rho NN}$ is the strong coupling constant, as
e.g. provided by the Bonn potential~\cite{mac87}. 

The axial vector type interaction has been suggested
by~\cite{sud68}. Unlike the $\rho$--type interaction the isospin
dependence is not restricted and may be isoscalar, --vector, and/or
--tensor type. The effective Lagrangian for the $a_1NN$ coupling for
example is given by
\begin{equation}
{\cal L}_{a_1NN} = g^T_{a_1NN}\;
\bar N\gamma_5 \sigma_{\mu\nu}
\frac{(p_{f}-p_{i})^\nu}{2m_N}\mbox{\boldmath$\tau$} N\;{\bf a}_1^\mu.
\label{eqn:ANN}
\end{equation}
Combined with the appropriate $T$--even vertex this leads to an
effective axial vector type exchange $NN$ potential~\cite{sud68},
\begin{equation}
V_{a_1}^T=i\;\frac{g^T_{a_1NN}g_{a_1NN}}
          {8m^2_N({\bf q}^2+m_{a_1}^2)}\;\ut_1\cdot\ut_2\;
          ( \us_1\cdot{\bf p}\;\us_2\cdot{\bf q}
           +\us_2\cdot{\bf p}\;\us_1\cdot{\bf q}
           -\us_1\cdot\us_2\;{\bf q}\cdot{\bf p}).
\end{equation}
The bounds on the $T$--odd coupling strengths arising from various
experiments have been discussed in ref.~\cite{her95}.  A more recent
bound not included there is from an improved analysis~\cite{eng95} of
the $^{57}$Fe $\gamma$--decay experiment~\cite{caltech}. Bounds are in
the order of 
10\% if derived from generic $T$--odd $P$--even observables and
slightly more than an order of magnitude smaller, if related to the
electric dipole moments~\cite{her95}.  To complete this section, note
that although possible, two boson exchanges have not been considered
up to now.

\section{Quark operators and effective NN interaction}

We now turn to effective $T$-odd $P$-even quark operators. The
simplest operator that leads to an effective $T$--odd $P$--even vector
type vertex $Vqq$ analogous to the $\rho$--type interaction
eq.~(\ref{eqn:VNN}) is
\begin{equation}
{\cal L}^T_{Vqq}
=g^T_{Vqq}\left(\bar u_1\gamma_\mu d_1 \bar d_2 \gamma^\mu u_2 -
\bar d_1\gamma_\mu u_1 \bar u_2 \gamma^\mu d_2\right).
\label{quarkV}
\end{equation}
Here $u$, $d$ denote flavored quark fields. Again the flavor
dependence is responsible for $C$--, viz. $T$--violation due to the
phase dependence. A tensor term has not been introduced for
simplicity. On the basis of eq.~(\ref{quarkV}) such a term will arise
in a natural way in the effective hadronic $\rho NN$ $T$-odd
Lagrangian through the quark structure effects as will be explained
below. The second generic quark operator utilizing axial vector
bilinear operators is given by~\cite{khr91},
\begin{equation}
\CL_{Aqq}^T=g^T_{Aqq}\;
\underbrace{\ovl{q}_1\gamma_5 \sigma_{\mu\nu}
\frac{(p_{f,1}-p_{i,1})^\nu}{2m_q}q_1}_
{\mbox{$\CO_1^T$}}\;\;
\underbrace{\ovl{q}_2i\gamma_5 \gamma^\mu q_2}_{\mbox{$\CO_2^E$}}
\quad + \quad (1\leftrightarrow 2)
\end{equation}
with the on--shell equivalent
\begin{equation}
\CO_1^T\;\CO_2^E = 
\ovl{q}_1i\gamma_5 \frac{(p_{f,1}+p_{i,1})_\mu}{2m_q}q_1
\;\;\bar q_2 i \gamma_5 \gamma^\mu q_2.
\end{equation}

In order to recover eqs.~(\ref{eqn:VNN}) and (\ref{eqn:ANN}) we
utilize the constituent quark model. This model has been rather
successful and valuable in reproducing gross features of low energy
phenomenae, such as mass spectra, form factors, coupling constants,
magnetic moments etc., see e.g.~\cite{quarks}.

To relate quark operators to effective hadronic operators we utilize 
the Fock space representation of hadrons in terms of
constituent quarks, viz.
\begin{equation}
~_{6q}\!\bramket{NN}{\CO^T\CO^E}{NN}_{6q}
\rightarrow\bramket{NN}{V^T_{MNN}}{NN}.
\label{fock}
\end{equation}
Since there is no low energy solution of QCD, the evaluation of the
matrix elements of the l.h.s of eq.~(\ref{fock}) needs further
consideration.  In general, the same problem arises in the context of
strong interactions. An extensive overview of the different approaches
to tackle the problem in this case has been given by
ref.~\cite{myh88}.  Here we follow the ideas first formulated in
ref.~\cite{web80}, and extensively studied for different quark models
in~\cite{boz83,virginia}. The resulting strong interaction potential
is a generic hybrid model connecting quark degrees of freedom with
effective meson nucleon degrees of freedom. The basic idea is
summarized in the following.

\begin{figure}[th]
\centering
  \leavevmode
  \epsfxsize=0.40\textwidth
  \epsffile{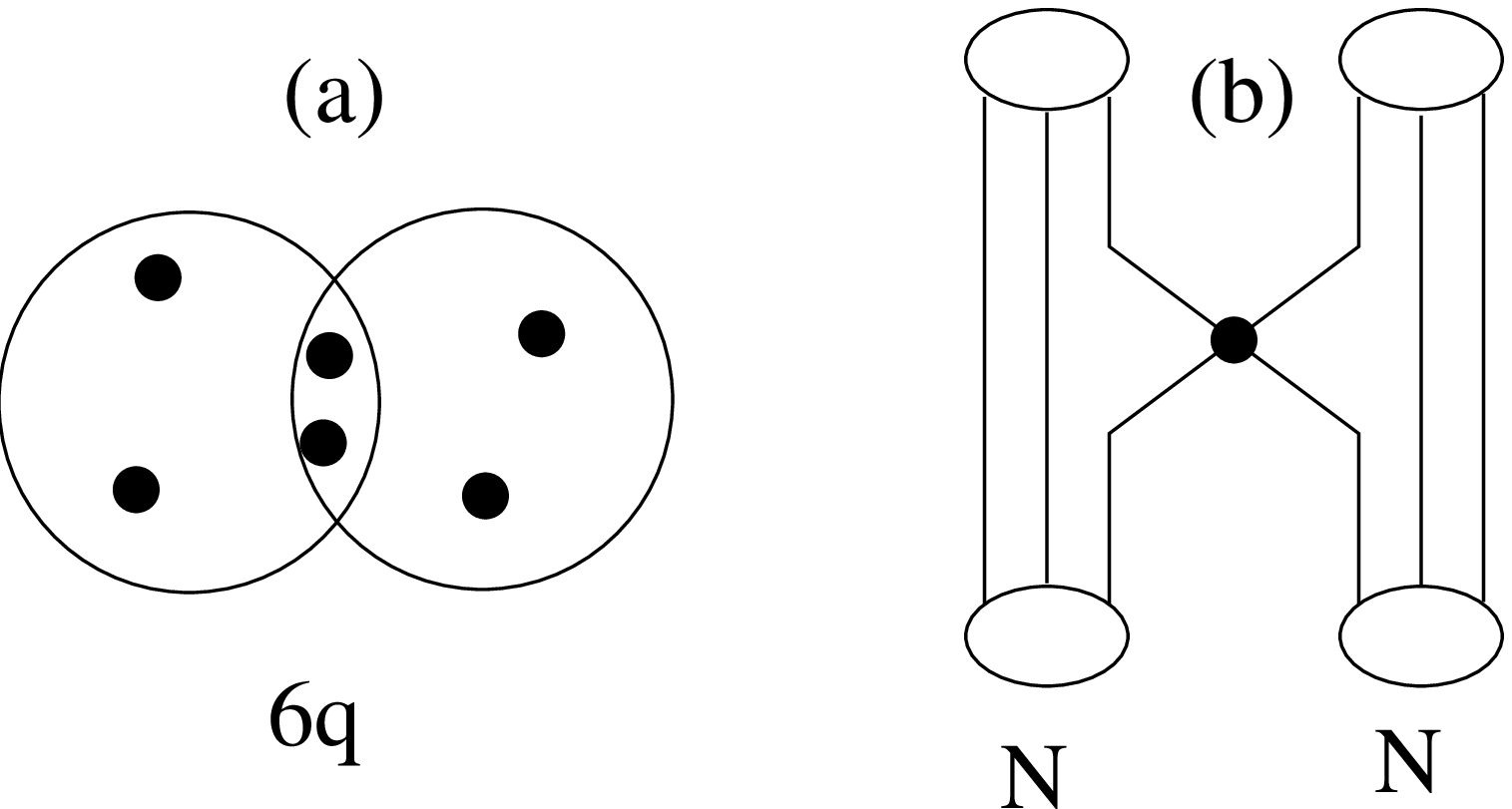}
\caption{\label{fig:qqNN} 
Pictorial demonstration of short range $T$--odd $NN$ interaction, a) $NN$
system as 6 valence quarks, b) factorization approximation.}
\end{figure}

Suppose the two nucleons overlap, and two quarks are sufficiently
close together. This situation is depicted in Figure~\ref{fig:qqNN}a).
Then, to begin with, the matrix elements may be evaluated without
introducing any mesonic fields. In terms of the constituent quark
model $q\bar q$ excitations are neglected (or partially parameterized
in the constituent quark mass).  Only at larger distances of the
nucleons, mesons are essential and may appear as $q\bar q$
correlations on the nonperturbative QCD vacuum~\cite{web80} that might
be the physical vacuum of the low energy regime~\cite{shi80,dia96}.
However, the appearance of mesons is disconnected from the problem of
$T$--odd force. Therefore, in the following we assume that the
hadronization  mechanism is the {\em same for both $T$--odd and the
  usual $T$--even strong interaction} and investigate the relative
strength of the $T$--odd matrix elements to the $T$--even matrix
elements.  This is done in the framework of the Virginia potential
that assumes a quark pairing mechanism to generate effective meson
nucleon coupling constants~\cite{web80,boz83,virginia}.  To illustrate
the quality of this ansatz Table~\ref{tab:virginia} shows the
resulting coupling constants using different quark models compared to
the values of a recent version of the Bonn
potential~\cite{mac87}.

In this framework we utilize the factorization approximation to
evaluate the matrix element of eq.~(\ref{fock}), see Figure~\ref{fig:qqNN}b)
\begin{equation}
\expect{\CO^T\CO^E}=
\expect{\CO^T}_{3q}\times\expect{\CO^E}_{3q}.
\label{quark}
\end{equation}
To demonstrate the calculation, we use the simple constituent quark
model. This model is supplemented by an explicit lower component,
which already occurs implicitly in the Dirac magnetic
moments~\cite{isgur} and in the two--body pair current through
electromagnetic gauge invariance~\cite{bey83,web85}. This way a
treatment of relativistic effects has been introduced, see
e.g.~\cite{bey85} and ref. therein.  The integration of the internal
degrees of freedom reads,
\begin{equation}
\expect{\CO}_{3q} = \int d^3\lambda d^3\rho \;\ovl{\psi}(\rho,\lambda) 
\CO \psi(\rho,\lambda)
\exp\left[{-i\sqrt{2/3}\;\mbf{q}\cdot\mbf{\lambda}}\right]
\end{equation}
with $\psi(\rho,\lambda)$ the three quark wave internal function. In
the rest system the space part of the wave function is given
by~\cite{boz83,web85,bey85}
\begin{equation}
\psi(\rho,\lambda) = N_0
\left(\frac{\alpha^2}{\sqrt{\pi}}\right)^{3/2}\;
\left(
\begin{array}{c}1\\\displaystyle 
-i\frac{\mbf{\nabla}_\lambda\cdot\mbf{\sigma}_3}{3m_q}
\end{array}\right)
\exp[-\alpha^2(\rho^2+\lambda^2)/2]
\end{equation}
where numerical values may be chosen as $\alpha\simeq m_q\simeq m_N/3 $, 
and $N_0^{-2}=(1+\alpha^2/4m_q^2)$.
The coordinates are normalized Lovelace coordinates,
viz. $\rho$ for the pair and $\lambda$ for the odd quark. Integration
is done in the Breit system.  
The symbol $\CO $ denotes either one of the vertices in (\ref{quark}).

Evaluation for the different type of operators leads to the following
expressions (using the isospin formalism for $u,d$ quarks)
\begin{eqnarray}
\langle i\gamma_5 \mbf{\tau}^I\rangle_{3q}&\rightarrow&
\left(\frac{5}{3}\right)^I N_0^2\;\frac{2m_N}{3m_q}
\langle i\gamma_5\mbf{\tau}^I_N \rangle_N\\
\langle i\gamma_5\gamma_\mu \mbf{\tau}^I\rangle_{3q}&\rightarrow&
\left(\frac{5}{3}\right)^I N_0^2\;(1-\frac{\alpha^2}{12m_q^2})
\langle i\gamma_5\gamma_\mu\mbf{\tau}^I_N \rangle_N
\label{axial}\\
\langle \gamma_\mu\mbf{\tau} \rangle_{3q}&\rightarrow&
\langle \gamma_\mu\mbf{\tau} \rangle_N
+ \left(\frac{10m_N}{9m_q} N_0^2-1\right)
\langle \sigma_{\mu\nu}
\frac{(p_{f,1}-p_{i,1})^\nu}{2m_N}\mbf{\tau}_N \rangle_N
\label{tensor}
\end{eqnarray}
The factors arising are related to the quark structure of nucleons.
Note, that in eq.~(\ref{axial}) one recognizes the well known coupling
of the axial vector current (for $I=1$), viz. $g_A=5/3$ of the
nonrelativistic constituent quark model, besides factors arising from
relativistic corrections due to the lower Dirac component. The latter
reduce the value of $g_A$ close to the experimental one~\cite{bey85}.
In eq.~(\ref{tensor}) a tensor coupling appears, which belongs to the
$\rho$ type $T$--odd exchange. Indeed due to the
quark structure factors its relative strength is larger than the first
term of the r.h.s. of eq.~(\ref{tensor}), and therefore preferable in
an ansatz of a $\rho$--type $T$--odd force as done  by
Simonius and Wyler~\cite{sim77}. The factor in front the
tensor term may be interpreted as ``anomalous'' coupling.  It appears
in analogy to the electromagnetic interaction, where the Pauli term of
the electromagnetic photon nucleon interaction can be recovered from a
pure Dirac coupling on quark level. This has been explained and shown
in ref.~\cite{boz83}.

\section{Results and Conclusion}

The resulting relation between quark and hadronic $T$--odd coupling
strength on the basis of the constituent quark model is,
for $\rho$ type exchange
\begin{equation}
g^T_{\rho NN} = \left(\frac{10m_N}{9m_q} N_0^2-1\right) g^T_{Vqq},
\end{equation}
and for axial type of exchange
\begin{equation}
g^T_{aNN} = \frac{2}{3}\frac{m_N}{m_q}
\left(1-\frac{\alpha^2}{12m_q^2}\right)^{-1}g^T_{Aqq}.
\end{equation}
Here the expression for isoscalar and isovector are the same. 

Similar results may be obtained using different types of quark
models. In the context of the Virginia potential those studied are the
MIT bag model and a relativistic model with linear confining
potential, see ref.~\cite{virginia}. These are used here in the same
way as demonstrated for the constituent quark model in the previous
section. The values for the quark structure effects evaluated
using typical quark model parameters
of low energy phenomenology are given in
table~\ref{tab:values}.  

In fact, due to the symmetries inherent in the quark pairing mechanism
(viz. the Virginia potential)
it is possible to arrive at the following relations between the
coupling constants, viz.
\begin{eqnarray}
g^T_{\rho NN}&=& (f_{\rho NN}/g_{\rho NN}) 
\;g^T_{Vqq}=\kappa\; g^T_{Vqq},\\
g^T_{aNN} &=& (g_{\pi NN}/g_{a_1 NN})\; g^T_{Aqq}
 =  (g_{\eta NN}/g_{f_1 NN})\; g^T_{Aqq}.
\end{eqnarray}
This equation shows that the factors appearing in the $\rho $ type
exchange may be related to the anomalous coupling $\kappa=f/g $. So,
inclusion of $\kappa$ might give a bound closer to the more basic
quark degrees of freedom.

In conclusion, provided the hadronization process does not
substantially differ for $T$--odd and $T$--even interactions, the
factors arising reflect the {\em nucleon} structure effects. The
origin of the structure factors are due to the spin, isospin structure
and the different mass scales (i.e. $m_q$ vs. $m_N$). These have also
been essential in deriving the relative strength of the strong
coupling constants as given in Table~\ref{tab:virginia}.

\section{Acknowledgment}
The author gratefully acknowledges support by the National Institute
for Nuclear Theory at the University of Washington, Seattle, during
his stay on the INT program ``Physics beyond the Standard Model at Low and
Intermediate Energies''.  This work has been supported by Deutsche
Forschungsgemeinschaft Be 1092/4-1.

\begin{table}
\caption{ \label{tab:virginia} Effective MNN coupling
  constants $g^2_{MNN}({\bf q}^2=0)/4\pi$ calculated from quark
  pairing mechanism~\protect{\cite{virginia}}. Only one overall
  quark meson coupling constant needs to be fitted.  For comparison 
  values of the Bonn potential have also been included. In
  brackets [f/g] is shown.}
$$
\normalsize
\begin{array}{ccclclclcl}
\hline\hline
\mbox{Meson}&(J^{PC},I^G)
&\multicolumn{2}{c}{\mbox{MIT}}
&\multicolumn{2}{c}{\mbox{lin. conf.}}
&\multicolumn{2}{c}{\mbox{CQM}}
&\multicolumn{2}{c}{\mbox{Bonn}}\\
\hline
f_0(1200)  &(0^{++},0^+)&4.1 &       &4.5 &       
&7.09 &       &&\\
a_0(960)   &(0^{++},1^-)&0.5 &       &0.5 &       
&0.79 &       &1.62&\\        
\eta(550)  &(0^{-+},0^+)&5.1 &       &5.32&       
&5.04 &       &&\\
\pi(138)   &(0^{-+},1^-)&14.0&       &14.8&       
&14.0 &       &14.08&\\
\omega(782)&(1^{--},0^-)&6.3 &[-0.4] &3.8 &[-0.5] 
&9.85 &[-0.49]&10.6 &[0.0]\\
\rho(763)  &(1^{--},1^-)&0.7 &[2.2]  &0.42&[3.2]  
&1.10 &[1.53] &0.41 &[6.1]\\
f_1(1285)  &(1^{++},0^+)&0.5 &[-1.5] &0.22& 
&0.59 &[0]    &&\\
a_1(119)   &(1^{++},1^-)&1.0 &[-1.5] &0.6 & 
&1.64 &[0]    &&\\
\hline\hline
\end{array}
$$
\end{table}

\begin{table}
\caption{ \label{tab:values} Factors arising from the quark structure
  of the $NN$ system, for the constituent quark model; (a)
  $m_q=0.33$GeV, (b) $m_q=0.22$GeV. }
$$
\begin{array}{lcccc}
\hline\hline
&\mbox{MIT}&\mbox{lin.}&\multicolumn{2}{c}{\mbox{CQM}}\\
&\mbox{bag}&\mbox{conf.}&(a)&(b)\\
\hline\\
g^T_{a_1NN}/g^T_{a_1qq}&3.5 & 5.0 & 2.9 & 4.9\\
g^T_{\rho NN}/g^T_{\rho qq}&2.2 &3.2&1.5&2.0\\
\hline\hline
\end{array}
$$
\end{table}

\begin{references}
\bibitem{kaon} J.H. Christenson, J.W. Cronin, V.L.Fitch, R.Tulay Phys.
  Rev. Lett. {\bf 13} 138 (1964)

\bibitem{wolf} L.  Wolfenstein, invited talk on ``Is there life after
  the Standard Model?'' INT Program on {\em Physics beyond the
    Standard Model} (1995)

\bibitem{her95} P. Herzceg, in {\em Symmetries and Fundamental
    Interactions in Nuclei} eds. W.C. Haxton, E.M. Henley, (World
  Scientific, Singapore, 1995) p. 89,

\bibitem{boehm95} F.  Boehm, in {\em Symmetries and Fundamental
    Interactions in Nuclei} eds. W.C. Haxton, E.M. Henley, (World
  Scientific, Singapore, 1995) p. 67; P. Herczeg, Hyp. Int. {\bf 43}
  (1988) 77

\bibitem{gould94} C.R. Gould, J.D. Bowman, Yu. P. Popov (eds.) {\em
    Time reversal invariance and parity violation} (World Scientific,
  Singapore, 1994)

\bibitem{her66}
 P. Herczeg, Nucl. Phys. {\bf 75} (1966) 655
\bibitem{bry71}
 R. Bryan, A. Gersten, Phys. Rev. Lett. {\bf 26} (1971) 1000, {\bf 27}
 (1971) 1102(E)
\bibitem{sud68}
 E.C.G. Sudarshan, Proc. Roy. Soc. {\bf A305}, (1968) 319
\bibitem{sim75}
 M. Simonius, Phys. Lett. {\bf 58B} (1975) 147
\bibitem{sim77}
 M. Simonius, D. Wyler, Nucl. Phys. {\bf A286} (1977) 182
\bibitem{hax83} W.C. Haxton, E.M. Henley, Phys. Rev. Lett. {\bf 51} (1983) 1937


\bibitem{her87}
 P. Herczeg, in {\em Tests of time reversal invariance
 in neutron physics}, eds, N.R. Roberson et al., (World Scientific
 Publishing, Singapore, 1987) p. 24.
\bibitem{ade85}
 E.G. Adelberger, W.C. Haxton Ann. Rev. Nucl. Part. Sci. {\bf 35}
 501 (1985).
\bibitem{tow94} I.S. Towner and A.C. Hayes, Phys. Rev. {\bf C 49},
  2391 (1994).
\bibitem{gud93} V.P. Gudkov, X.-G. He, and B.H.J. McKellar, Phys. Rev. {\bf
      C47}, 2365 (1993). 
\bibitem{bey89}
 M. Beyer, Nucl. Phys. {\bf A493} 335 (1989).
\bibitem{bey93}
 M. Beyer, Phys. Rev. {\bf C48}  906 (1993).
\bibitem{hax93} W.C. Haxton and A. Hoering, Nucl. Phys. {\bf A 560}, 469
  (1993). 
\bibitem{mus94} W.C. Haxton, A. H\"oring, M.J. Musolf, Phys. Rev. D50
  3422 (1994).
\bibitem{cre79}
 R.J. Crewther, P. di Vecchia, G. Veneziano, E. Witten, Phys. Lett. {\bf
 88B} 123 (1979).
\bibitem{alt92}
 I.S. Altarev et al., Phys. Lett. {\bf B276} 242 (1992).
\bibitem{smi90}
 K.F. Smith, Phys. Lett. {\bf B234} 191 (1990).
\bibitem{pen92}
 J.M. Pendlebury, Nucl. Phys. {\bf A546} 359c (1992).
\bibitem{mon94}
 Montanet et al. (Particle Data Group), Phys.Rev. {\bf D50} 1173 (1994).
\bibitem{her95a} P. Herczeg, private communication: P. Herzceg,
  J. Kambor, M. Simonius, and D. Wyler to be published.
\bibitem{eng96}J. Engel, P. Frampton, and R. Springer, preprint
  nucl-th/9505026  (1995).
\bibitem{zuber} C. Itzykson and J.-B. Zuber, {\em Quantum field
    theory} (McGraw Hill International, Singapore, 1980).
\bibitem{herINT} P. Herczeg, private communication.
\bibitem{eng95}  M.T. Ressell, J. Engel, P. Vogel, preprint MAP-189,
  nucl-th/9512013 (1995).
\bibitem{caltech}
 N.K. Cheung, H.E. Henrikson, F. Boehm, Phys. Rev. {\bf C16}, 2381 (1977).
\bibitem{khr91}
 I.B. Khriplovich, Nucl. Phys. {\bf B352} 382 (1991).
\bibitem{quarks} see e.g. F.E. Close, {\em An introduction to quarks and
    partons} (Academic Press, 1979); A. Le Yaouanc, L. Oliver, O.
  Pene, J.-C. Raynal, {\em Hadron transitions in the quark model}
  (Gordon and Breach, 1988); R.K. Bhaduri, {\em Models of the nucleon:
    from quarks to soliton}  (Addison-Wesley, 1988, Lecture notes and
  supplements in physics, 22).  
\bibitem{myh88} F. Myhrer, Rev. Mod. Phys. 60 629 (1988)
\bibitem{web80} H.J. Weber, Z. Phys. {\bf A297} (1980) 261, {\bf
    A301}, 141 (1981), Phys. Rev. {\bf C26}, 2333 (1982) 
\bibitem{boz83} M. Bozoian and H.J. Weber, Phys. Rev. {\bf C28} 811
  (1983) 
\bibitem{virginia} M. Beyer, H.J. Weber, Phys. Lett. {\bf
    146B} (1984) 383; Phys. Rev.  {\bf C35} (1987) 14
\bibitem{shi80} M.A. Shifman, A.I. Vainshtein, and V.I. Zakharov,
  Nucl. Phys. {\bf B147}, 385 and 448 (1980)
\bibitem{dia96} for a recent discussion of constituent quarks see:
  D. Diakonov, Gatchina preprint, nucl-th/960323.
\bibitem{mac87} R. Machleidt, K. Holinde and Ch. Elster, 
Phys. Rep. {\bf 149}, 1 (1987)
\bibitem{isgur} N. Isgur and G. Karl, Phys. Rev. {\bf D18}, 4187
  (1978); {\bf D 19}, 2653 (1979).
\bibitem{bey83} M. Beyer, D. Drechsel, and M.M. Giannini, 
       Phys. Lett. {\bf 122 B}, 1 (1983)
\bibitem{web85} H.J. Weber, M. Weyrauch Phys. Rev. {\bf C32}, 1342
  (1985) 
\bibitem{bey85} M. Beyer and S.K. Singh, Phys. Lett. {\bf B 160},
  26 (1985); Z. Phys. {\bf C31}, 421 (1986)
\end{references}
\end{document}